\magnification\magstep 1
\baselineskip=0,59 true cm
\vsize=21 true cm
\topinsert\vskip 1 true cm
\endinsert
{\centerline{\bf {MODULAR GROUPS, VISIBILITY DIAGRAM AND QUANTUM
HALL EFFECT}}}
\vskip 2 true cm
{\centerline{\bf{Yvon Georgelin$^a$, Thierry Masson $^b$ and 
Jean-Christophe Wallet$^a$}}}
\vskip 0.5 true cm
{\centerline{$^a$ Division de Physique Th\'eorique{\footnote
\dag{\sevenrm{unit\'e de Recherche des universit\'es Paris 11 et Paris 6
associ\'ee au CNRS}}}, Institut de Physique Nucl\'eaire}}
{\centerline{F-91406 ORSAY Cedex, France}}
\vskip 0.5 true cm
{\centerline{$^b$ Laboratoire de Physique Th\'eorique et
Hautes \'Energies{\footnote
*{\sevenrm{Laboratoire associ\'e au CNRS, URA D0063}}},}}
{\centerline{Universit\'e de Paris-Sud, B\^at 211}
{\centerline{F-91405, Orsay Cedex, France}}
\vskip 2 true cm
{\bf{Abstract:}} \par
We consider the action of the modular group $\Gamma (2)$ on the set of 
positive rational fractions. From this, we derive a model for a classification
of fractional (as well as integer) Hall states which can be visualized on two
``visibility" diagrams, the first one being associated with even denominator
fractions whereas the second one is linked to odd denominator fractions. We use
this model to predict, among some interesting physical quantities, the 
relative ratios of the width of the different
transversal resistivity plateaus. A numerical simulation of the tranversal
resistivity plot based on this last prediction fits well with the present 
experimental data.
\vskip 1 true cm
{\noindent{IPNO}}-TH-96-27 (december 1996)\hfill\break
{\noindent{LPTHE} Orsay 96/54\hfill\break
\vfill\eject
{\bf {I) Introduction}}\hfill\break
\vskip 0,2 true cm
The quantum Hall effect has been an intensive field of theoretical and
experimental investigations, since the discovery of the two dimensional
quantized Hall conductivity by [1] for the integer plateaus and by [2] for the
fractional ones.\par
The pioneering theoretical contributions [3a-c] relating the basic 
features of the
hierarchy of the Hall plateaus with the properties of a two dimensional
incompressible fluid with fractional charges collective excitations [3a] have
triggered further works dealing with condensed matter theory [4], quantum field
theory [5] and mathematical physics [6]. It appears now that the two dimensional
electron system in the quantum Hall regime is associated with a complicated
phase diagram with a lot of possible transitions between the various phases,
whose better understanding deserves further experimental and theoretical 
studies.\par
As far as the theoretical viewpoint is concerned, two important ingredients
underlying numerous works can be singled out: i) The flux attachement
transformation relating quantum Hall states [7] for which the Chern-Simons
theory is a suitable field theory framework; ii) the hierarchical structure
of the set of Hall states [8]{\footnote\dag{\sevenrm{given a basis of Laughlin
states, one can obtain elements of the set of Hall states through condensation
of quasiparticles of other elements of the set}}} where particle-hole symmetry
play an important role. The combination of the above ingredients
into a Chern-Simons-Landau-Ginburg theory has led the authors of [9] to propose
a topology of the phase diagram (in the external magnetic field-disorder plane)
together with an attempt to explain the possible origin of superuniversality
ruling the considered transitions.\par
It appears that (at least some of) the above ideas can be formulated in term of
modular transformations, that is, discrete symmetry transformations acting on
some parameter space of the quantum Hall system and pertaining to the modular
group or one of its subgroups. Modular transformations in the framework of
quantum Hall effect have been considered from different viewpoints [10].\par
In a recent work [11], two of us have constructed a classification of the 
quantum Hall
states based on arithmetical properties of the group
$\Gamma (2)$ of modular transformations acting on the three cusps
$\{i \infty \}$, $\{0\}$ and $\{1\}$ of its fundamental domain in the
Poincar\'e half-plane. The proposed classification reproduces and refines
the Jain hierarchical one [12]. In particular, it has been
obtained families of quantum Hall fluid states
(plus the insulator state),
each family being generated from a metallic state labelled
by an even denominator fraction $\lambda = {{(2s+1)}/2r}$ .\par
The purpose of the present paper is to extract more algebraic
properties linked with
this classification, then to translate them into a graphical representation, 
called the visibility diagram [13] which generates
Farey sequences [14] together with an associated diagram called the dual 
diagram.  From this analysis, we predict a global organization of 
the various Hall conductivity states stemming from the 
action of $\Gamma(2)$ as well as some 
important quantitative features of the Hall fractional
conductivity plateaus such as the relative ratios of their
width (at least for not a to strong applied magnetic field). Furthermore, we
obtain from a numerical construction a transverse resistivity 
plot which fits very well with the experimental data.\par
The paper is organized as follows. In section II we present the usefull
material underlying our classification of the quantum Hall states. In section
III, we introduce the visibility diagram together with its associated dual
diagram in connection with the quantum Hall
effect. The corresponding physical information encoded in these diagrams is
discussed in section IV where we also confront a numerical simulation of the
resistivity plot stemming
from our construction to the experimental data. Finally, we summarize our
results and conclude in section V.\par
\vskip 0,5 true cm
{\bf{II) The group $\Gamma(2)$ and the quantum Hall states}}
\vskip 0,2
 true cm
In this section, we summarize the essential tools that will be needed in
the sequel as well as some of the main results of the previous work [11] to
which we refer for more details.\par
Let ${\cal{P}}$ be the upper complex plane and $z$ a complex coordinate on
${\cal{P}}$ (Im$z>0$). We recall the inhomogeneous modular group
$\Gamma(1)$ ($=$PSL$(2,Z)$) is the set of transformations $G$ acting on 
${\cal{P}}$ and defined by:
$$G(z)={{az+b}\over{cz+d}};\ a,\ b,\ c,\ d\in Z;\qquad 
ad-bc=1\quad ({\hbox{unimodularity condition)}}  \eqno(2.1).$$
The modular group $\Gamma(1)$ has two generators given by:
$$T(z)=z+1;\qquad S(z)=-{{1}\over{z}}  \eqno(2.2).$$
The group $\Gamma(2)$ that will be the building block of the following
construction is a subgroup of $\Gamma(1)$. $\Gamma(2)$ is known in the
mathematical litterature as the principal congruence unimodular group at level
2. It is the set of transformations $G$ acting on ${\cal{P}}$ defined by
$${\hbox{eqn. (2.1)}};\qquad a\ {\hbox{and}}\ d\ {\hbox{odd}},\ b\ 
{\hbox{and}}\ c\ {\hbox{even}}
  \eqno(2.3).$$
$\Gamma(2)$ is a free group generated by
$$T^2(z)=z+2;\qquad \Sigma(z)=ST^{-2}S(z)={{z}\over{2z+1}}  \eqno(2.4).$$
For more mathematical properties of $\Gamma(2)$, see e.g [15]. Here, we notice
that $\Gamma(2)$ is a common subgroup of two other subgroups of $\Gamma(1)$
which have been already considered from different viewpoint in connection to
the quantum Hall effect. The first one, $\Gamma_S(2)$, is generated by $T^2$
and $S$ and is the natural symmetry group of the Landau problem on the torus;
it appears in the construction of the many body Landau states on the torus [16]
in term of the Coulomb gaz vertex operators [17]. The second subgroup of
$\Gamma(1)$, $\Gamma_T(2)$, which contains $\Gamma(2)$, is related to the
similarity tranformations used by Jain in his 
hierarchical model [12]. Recall the $\Gamma_T(2)$ is generated 
by $T$ and $\Sigma$.\par
The tranformations $G\in\Gamma(2)$ can be written as
$$G(z) = {{(2s+1)z + 2n}\over {2rz + (2k+1)}} \eqno(2.5),$$
where $k$, $n$, $r$, $s\in Z$ satisfy the unimodularity
condition
$$ \qquad (2s+1)(2k+1)-4rn=1 \eqno(2.6).$$
Now, identify any fraction $z=p/q$ with a filling factor and select a given Hall
metallic state labelled by $ \lambda = {(2s+1)\over 2r}$ with
$r\ge0,s\ge0$. As shown in [11],
there is a family of
transformations $G^{\lambda}_{n,k}\in\Gamma(2)$
(with $n$ and $k$ still satisfying (2.6))
which send $z=i\infty$ onto $\lambda$. Then, from the images
$G^{\lambda}_{n,k}(0)$ and $G^{\lambda}_{n,k}(1)$ of
$0$ and $1$ by the tranformations $G^\lambda_{n,k}$, one obtains
a (Jain-like) hierarchy of Hall states surrounding the metallic
state $\lambda$ on the resistivity plots.\par
For example, the double Jain family $J^i_{1/2}$ of states surrounding the
metallic state $\lambda = {1\over 2}$
$$ J^+_{1/2} = 1/3,\quad 2/5,\quad 3/7,\qquad \cdots {N/(2N+1)}\eqno(2.7a),$$
$$ J^-_{1/2} = 2/3,\quad 3/5,\quad 4/7,\qquad \cdots {N/(2N-1)}\eqno(2.7b)$$
can be easily recovered in our scheme from the images  $G^{1/2}_{n,k}(0)$ and
$G^{1/2}_{n,k}(1)$ with respectively $n\ge0$ for (2.7a) and
$n<0$ for (2.7b). Notice that the construction separates the even denominator
Hall fractions from the odd denominator one so that it may be possible to take
into account a possible particle-hole symmetry within the present scheme. Other
families surrounding any metallic state (fraction with even denominator) can be
constructed in the same way so that all the experimentaly observed Hall states
can be taken into account in the present construction. This will be discussed
further in section IV.\par
In the next section, we will be concerned with a diagrammatic description of
the classification of the quantum Hall states which permits one to visualize
the global organization of the various Hall conductivity states stemming from
the action of $\Gamma(2)$. To reach this goal, we first introduce a convenient
parametrization of the transformations $G^{\lambda}_{n,k}$. Indeed, it can be
easily verified that the $G^{\lambda}_{n,k}$'s, for fixed
$\lambda ={(2s+1)/2r}$, can be recast into the form
$$ G^{\lambda}_m(z)= {{(2s+1)z+2m(2s+1)+2b}\over {2rz+2m(2r)+2d+1}} \eqno(2.8)$$
where $m$ is any integer and $b$ and $d$ are also arbitrary
integers satisfying the unimodularity condition
$$(2s+1)(2d+1)-4rb=1 \eqno(2.9).$$
For a given $\lambda$, fixing $b$, $d$ and varying $m$
from $-\infty$ to $\infty$, a complete hierarchy of states is recovered as
before [11] from
$G^{\lambda}_m(0)$ and $G^{\lambda}_m(1)$. Furthermore, one has
$$\lim_{m\to \pm \infty}G^{\lambda}_m(z)= {\lambda} \eqno(2.10),$$
which reflects the fact that the sequence of Hall fluid fractions
converges to the
fraction $\lambda$ labelling a given metallic state which defines
that sequence. This point has to be compared to the general topology of the
recently proposed phase diagrams for the quantum Hall effect where, roughly, 
each metallic state (even denominator state) appears to be surrounded by a
definite family of Hall liquid states (odd denominator states). We will discuss
this point in section IV.\par
From the above parametrization (2.8), we now construct a shift operator 
$A^{\lambda}\in\Gamma(2)$ corresponding to each metallic state $\lambda$, 
which acts on the fluid states (and possibly on the
 insulator state $\nu = 0$). We define this
shift operator $A^{\lambda}$ (which depends only on $\lambda$) as 
the following operator product
$$A^{\lambda} = G^{\lambda}_{m+1}\circ (G^{\lambda}_m)^{-1} \eqno(2.11),$$
whose action is given by
$$A^{\lambda}(z)={{(1-2(2r)(2s+1))z+2(2s+1)^2}\over
    {-2(2r)^2z+1+ 2(2s+1)2r}} \eqno(2.12).$$
The action of $A^{\lambda}$ to a given rational
fraction $p/q$ equals to $G^{\lambda}_m(0)$ or
$G^{\lambda}_m(1)$
gives another rational fraction equals respectively to $G^{\lambda}_{m+1}(0)$
or $G^{\lambda}_{m+1}(1)$, so that $A^\lambda$ acts as a shift
operator, raising the number $m$ by one unit. The inverse
operator $(A^{\lambda})^{-1}$
performs the corresponding lowering in $m$.\par
The operators $A^{\lambda}$ satisfy the following relations
$$A^{\lambda}(z)=z \qquad \iff z= \lambda \eqno(2.13),$$
$$Tr[A^{\lambda}]=2 \eqno(2.14), $$
$$det[A^{\lambda}]=1 \eqno(2.15).$$
where $[A^{\lambda}]$ denotes the two by two matrix corresponding
to the modular transformation $A^{\lambda}$. Eqns. (2.14) and (2.15) are just a
mere consequence of the definition of the group $\Gamma(2)$. Notice that (2.14)
reflects the parabolic
nature of the transformations $A^{\lambda}$ which means that $A^{\lambda}$
may have no more than two fixed points (which must be necessarily real [15]).
Therefore, eqn.(2.13) means that the metallic state ${\lambda}= {(2s+1)}/2r$ is
actually the unique fixed point for the shift operator $A^\lambda$. This last
property may be extended to the case where $r=0$, which corresponds to
the point at infinity on the
real axis.
\vskip 0,5 true cm
{\bf {III) Visibility diagram and dual diagram.}}
\vskip 0,2 true cm
We are now in position to introduce a diagrammatic description that
we find particularly usefull and predictive for the
classification of the quantum Hall states. This diagrammatic
tool, which permits one to obtain quantitative results on the different
quantum Hall states is called the ``visibility diagram" and has been already
used in other areas of physics [13]. We will also introduce another diagram,
hereafter called the dual diagram, which will allow us to extract in a
straighforward way some of the information encoded in the visibility 
diagram.\par
\vskip 0,2 true cm
{\bf{IIIa) The visibility diagram.}}\par
This diagram is a simple pictorial device
for the construction of Farey sequences [14] of rational numbers, each one
being associated to a given vertex of a two dimensional
square lattice. \par
Let us consider the two dimensional square lattice depicted on fig. 1 where the
vertices are indexed by a couple of positive or nul relatively prime 
integers $(q,p)$. Since the action of $\Gamma(2)$ preserves the even 
or odd nature of $q$ and since a special role
is played by the even denominator fractions in the
sequel, we find convenient to use different symbols (see fig. 1) to
separate the vertices belonging to a row with $q$ even from those belonging to
a $q$ odd row. Therefore, the visibility diagram is made of the vertices of 
couples $(q,p)$ of positive integers which are not hidden
by any other vertex for an observer located at the origin
$(0,0)$. Notice that the origin is not considered as a visible point.\par
It is known in arithmetics that all the Farey ordered
sequences $F_n, (n=1,2,\ldots )$ of rational
fractions $p/q$ with $p$ positive and $ 0\le q\le n$
are in one to one correspondance with the visible
vertices $(q,p), 0\le q\le n$. Strictly
speaking, for any Farey sequence $F_n$, $p$ and $q$ are required to
satisfy $p\le n$ and $q\ne 0$; here, since we have in mind to apply the above
machinery to the fractional quantum Hall effect,
we relax somehow the above requirement, thus allowing fractions
with $p>n$ (and formally the ``infinite fraction"
$\infty \equiv {1/0} \iff (0,1)$).\par
Corresponding to this identification, the group $\Gamma (2)$
acts on the vertices of the visibility diagram
in a rather simple way: it is actually represented by the linear
action of two by two matrices on the two components $(q,p)$ of
the vertices, a property we will use in the following. \par
Now, it is easy to draw on the visibility
diagram each family of vertices corresponding
to the previously introduced fractions
$G^{\lambda}_m(0)$ and $G^{\lambda}_m(1)$. \par
Let us describe in details what kind
of picture we obtain (see fig.2). First of all,
for a given $\lambda = {(2s+1)/2}$, $r=1$,
the quantum Hall rationals numbers $G^{\lambda }_m(0)$ and
$G^{\lambda }_m(1)$, where $m$ is any integer, are all located on each
parallel sides of distinct unbounded bevelled left
end side stripes. The two sides of each stripe start from the vertices
$(1,s)$ and $(1,s+1)$ and their slope is equal to $\lambda = {(2s+1)/2}$, as it
can be easily seen by using for instance (2.8) and (2.9) from which one obtains
after some algebra the equations defining the two parallel sides of the
stripes. They are given in the $(q,p)$ plane by
$P=({{2s+1}\over{2}})Q\mp{{1}\over{2}}$ where the minus (resp. plus) sign
refers to the side starting from the vertex $(1,s)$ (resp.($1,s+1)$). This 
agrees with the generation mechanism
we gave previously for each Jain type family [11]; recall that,
as far as the action of $\Gamma (2)$ is concerned,
the couple of cusps $( \{s\}, \{s+1\}) $ may be substituted to
the equivalent couple $( \{0\}, \{1\}) $. The metallic state
$\lambda$ itself is shown to be surrounded on the visibility
diagram by the family of fluid states (and eventually by the
insulator whenever $\lambda = 1/2$) it generates. We will
call such families $(\lambda = (2s + 1)/2)$ the principal
stripes ($r=1$).\par
Glued on each side of any principal stripe there is an infinite
number of other secondary, tertiary... stripes ($r=2,3...$). They are
all similar in shape to a principal one: Two infinite parallel
sides, a bevelled left end side and each one labelled
by a metallic state $\lambda = {(2s+1)/2r}$. The states
$G^{\lambda}_m(0)$ and $G^{\lambda}_m(1)$, with $m$
any integer, are all located on the parallel sides of
the stripe $\lambda$ in such a way that each metallic
state find itself isolated from the other metallic states
by the members of its family on a stripe of slope
$\lambda$. Starting from the secundary stripes ($r=2$), it is
possible to construct families of tertiary stripes,
quartenary stripes... and so on, families of stripes
of any order. The equations defining the two sides of a stripe of order $r$ are
easily derived from (2.8) and (2.9). They are given in the $(q,p)$ plane 
by $P=({{2s+1}\over{2r}})Q\mp{{1}\over{2r}}$. A pecularity appears
on the diagram concerning the
Hall insulator state $(\nu =0)$. Its vertex representation
is $(1,0)$; clearly it pertains to a given family $\lambda$,
only for $\lambda = {1/2r} (r\ge 1)$. This last property could
have some importance for a possible direct phase transition
between that state and another state.
The reader has certainly noticed that the quadrant
$ p\ge 0,q\ge 0$ is not completely
covered by stripes, a (half) stripe shaped region
is left uncovered, it is defined by $p\ge 0, 0\le q \le 1$.
We shall also  come back on this last pecularity.\par
On any fraction of any given stripe
$\lambda$ acts the step operator $A^{\lambda}$ or its inverse in
such a way that the metallic state $\lambda = {(2s+1)/2r}$
located at the lattice vertex $(2r,2s+1)$ is a fixed
point for $A^{\lambda}$. The fluid states (and eventually the
insulator state) on the edge of the corresponding stripe
are all transformed among themselves under that action,
any of them is in fact the image of a $G^{\lambda}_m(0)$
or a $G^{\lambda}_m(1)$ under some power of $A^{\lambda}$ and its inverse. In
fact, all the information encoded in this diagram can be thoroughly reproduced
by using the $A^\lambda$ and $(A^\lambda)^{-1}$'s. This can be rephased by
saying that the $A^\lambda$'s are the algebraic counterpart of the graphical
representation of the (tree-like) structures formed by the Hall states on 
the above diagram.\par
\vskip 0,2 true cm
{\bf{IIIb) The dual diagram}}\par
Observe that the previous structure on the visibility diagram is a mere
consequence of a well-known theorem in arithmetics which states that, for any
given two relatively prime positive integers $p$ and $q$, there exist two other
(necessarily relatively prime) positive integers $a$ and $b$ satisfying 
$pb-qa=1$ or $pb-qa=-1$. Using this theorem, it is easy now to realize that 
the set of $(a,b)$ pairs stemming from these later equations, for fixed
$(p,q)$
and $q$ even, are all located on the two sides of the stripe labelled by
$\lambda={{p}\over{q}}$ depicted on fig.2.\par
But one could instead consider $q$ odd so that another structure can be
obtained on the visibility diagram where now stripes labelled by
$\lambda={{p}\over{q}}$ with $q$ odd can be defined as shown on fig.3. We will
call this resulting structure the dual diagram. Indeed, it can be easily seen
that any stripe on fig.2 corresponds to a given direction on the dual diagram
(depicted by dotted lines on fig.3), and conversely. This diagram will prove
usefull to evaluate the relative ratios of the width of the Hall plateaus.\par
\vskip 0,5 true cm
{\bf {IV) Physical discussion.}}
\vskip 0,2 true cm
We are now in position to extract some physical information from the two
diagrams that have been defined. We first identify any fraction $p/q$ with the
filling factor $\nu$ (what we implicitely assumed in the previous sections).
This picture suggests that the $\Gamma(2)$ symmetry transformations act on some
parameter space describing the quantum Hall system. Recall that the 
metallic (resp. liquid) states are labelled by fractions with
even (resp. odd) denominator.\par
First of all, keeping in mind the above identification, it is obvious that our
construction treats on an equal footing integer and fractional states. Now,
consider the structure given on fig. 2. One observes that each principal stripe
bears self-similar branched structure built from that stripe and all its
descendants. Notice that the resulting global organization of the states in the
stripes agrees quite well with the experimentaly observed one. We point out 
that our construction predicts the existence of
an infinite number of families with self-similar (tree-like) structure. As far
as the present experimental situation is concerned, (see [18]), the Jain
family $\lambda=1/2$ corresponding on fig.2 to the stripe
with the largest width, has the most numerous well established observed states
($N_{obs}\approx 13$). The other observed families labelled respectively by
$\lambda=3/2,1/4,3/4,5/2,5/4$ and corresponding respectively to stripes of
decreasing width are associated to a decreasing number of observed states
(respectively $N_{obs}\approx7,6,5,4$). Roughly speaking, this suggests that
the width of a stripe may be related to the (experimental) difficulty to
observe the associated states. Notice that the insulator state $\nu=0$($=0/1$
in our picture) does not label any stripe.\par
Now consider the action of the two generators of $\Gamma(2)$ given in (2.4) 
on the visibility diagram. $T^2$ can be interpreted as a Landau shift operator
; $\Sigma$ corresponds to a flux attachment operator. On the
visibility diagram, the action of $T^2$ on a vertex $(q,p)$ gives rise to 
a vertical shift $(q,p)\to(q,p+2q)$ whereas the action of $\Sigma$ on $(q,p)$ 
produces a horizontal shift $(q,p)\to(q+2p,p)$. Since flux attachement can be
obtained by magnetic field variation, a variation of the magnetic field
is associated on the visibility diagram with a corresponding horizontal shift.
Observe now that the stripes of the dual diagram are labelled by fractions with
odd denominator, each of which corresponding to a plateau in the Hall
conductivity. Putting all together, this suggests to identify (up to an overall
dimensionfull factor) the ``horizontal" width of any stripe of the dual
diagram defined by the intercept of any horizontal line with that stripe with
the width of the corresponding plateau. Notice that both integer and fractional
plateaus are involved in this picture as already mentioned at the beginning of
this section.\par
It is a straighforward computation to show that the ``horizontal" width of any
stripe labelled by $\lambda=p/q$ is equal to $2/p$. Then, the above
identification leads to a prediction of
all the ratios of the widths of the Hall plateaus given by
$${{\gamma(\nu_1=p_1/q_1)}\over{\gamma(\nu_2=p_2/q_2)}}={{p_2}\over{p_1}} 
\eqno(4.1)$$
where $\gamma(\nu)$ denotes the width of the plateau labelled by $\nu$ and can 
be expressed as
$$\gamma(\nu=p/q)={{2}\over{p}}\gamma_0  \eqno(4.2),$$
where the overall parameter $\gamma_0$ has dimension of a magnetic field. At 
this point, one
important remark is in order. It is well known that the width of the plateaus
decreases when the temperature increases so that for sufficiently hight
temperature the classical behavior for the Hall resistivity is recovered.
Therefore, if the above identification is correct, eqn.(4.1) should be valid
only at zero temperature.\par
Nevertheless, one might already expect to obtain a reasonably good agreement
with the present (non zero temperature) experimental data. In order to check
this, we have confronted the prediction (4.1) with an experimental resistivity
plot [19]. To do this, we have to fix two parameters. The first one is 
the width of a {\it{given}} transverse resistivity 
plateau which will determine the value of $\gamma_0$. In
the present numerical analysis, we choose to fit this value with
the $\nu=3/2$ experimental plateau. The second parameter is
the position of the center of {\it{this choosen}} plateau on the to be
determined resistivity plot. We assume that this plateau is centered around its
corresponding filling factor value $\nu=3/2$. These two requirements will
determine completely all the widths and positions on the other plateaus. The 
numerical result is shown on fig.4 and exhibits a good agreement with
the corresponding experimental plot (fig.5). \par
\vskip 0,5 true cm
{\bf {V) Conclusion.}}
\vskip 0,2 true cm
In this paper we have exploited some arithmetic properties linked with the
classification of the quantum Hall states stemming from the action of
$\Gamma(2)$ . These properties can be well synthetized in a graphical
representation based on a visibility diagram, in which the resulting global 
organization of all the quantum Hall states is found to form (tree-like)
self-similar structures. In particular, this structure is in complete agreement
with the (experimental) hierarchy of observed states.\par
Furthermore, we used some of the arithmetic properties rooted in this
construction together with a physical interpretation of the two generators of
$\Gamma(2)$ to conjecture the (zero temperature) ratios of the widths of 
the Hall plateaus. Moreover, this suggests in particular that the zero
temperature values of these ratios are universal. We tried to confront the
predictions obtained from this conjecture to the present (non zero temperature)
experimental data by constructing numerically the corresponding resistivity
plot. This latter is in good agreement with some experimental resistivity
plot (see on fig.5).\par
As it has been already mentionned in this paper, the $\Gamma(2)$
transformations could be understood as an infinite set of discrete 
transformations acting on some parameter space of the quantum Hall system.
Therefore, it should be interesting to use this $\Gamma(2)$ symmetry to
construct a model for the quantum Hall effect, for instance in the spirit of
ref. [9] and first of ref. [10]. On the other hand, it should be also
interesting to incorporate the $\Gamma(2)$ symmetry into a renormalization
group framework  to study the possible relations it induces among the 
fixed points that might provide some new insight on the possible origin of the
superuniversality property shared by the observed transitions between Hall
states. These questions are presently under study.\par
\vskip 3 true cm
{\bf {Aknowledgments}}\par
We are indebted to A. Comtet, M. Dubois-Violette, J. Mourad 
and S. Ouvry for valuable discussions.
\vfill\eject
\topinsert\vskip 2 true cm
\endinsert
{\bf {REFERENCES}}\par
\vskip 1 true cm
\item {[1]:} K. von Klitzing, G. Dorda and M. Pepper, Phys. Rev. Lett. 45 (1980)
494.
\item {[2]:} D. C. Tsui, H. L. St\"ormer and A. C. Gossard, Phys. Rev. Lett. 48
(1982) 1559.
\item {[3a]:} R. B. Laughlin, Phys. Rev. Lett. 50 (1983) 1395.
\item {[3b]:} F. D. M. Haldane, Phys. Rev. Lett. 51 (1983) 605.
\item {[3c]:} B. I. Halperin, Phys. Rev. Lett. 52 (1984) 1583.
\item {[4]:} For a review, see, e.g. Quantum Hall effect, by M. Stone, World
Scientific (Singapore) 1992.
\item {[5]:} We have in mind models related to Chern-Simons theory, Coulomb gaz
approach and rational conformal field theories; see [4] and references therein.
\item {[6]:} A. Connes, in ``G\'eom\'etrie non commutative, InterEditions,
Paris (1990); for a review see J.Bellissard, A. van Elst and H. Schulz-Baldes,
J. Math. Phys. 35 (1994) 5373.
\item {[7]:} S.M. Girvin and A.H. MacDonald, Phys. Rev. Lett. 1252 (1987) 58;
N. Read, Phys. Rev. Lett. 62 (1989) 86; A. Lopez and E. Fradkin, 
Phys. Rev. B44 (1991) 5246; S.C. Zhang, T. H. Hanson and 
S. Kivelson, Phys. Rev. Lett. 82 (1989) 62.
\item {[8]:} see ref. [3b] and [3c]; J.K. Jain, Phys. Rev. Lett. 63 (1989) 199.
\item {[9]:} S. Kivelson, Dung-Hai Lee and Shou-Cheng Zhang,
Phys. Rev. B46 (1992) 2223.
\item {[10]:} see e.g. E. Fradkin and S. Kivelson, Nucl. Phys. B474 (1996)
 543; C. P. Burgess and C. A. L\"utken, MacGill preprint,
McGill-96/39 (1996) (cond-mat/9611118); L. P. Pryadko and 
S. C. Zhang, cond-mat/9511140.
\item {[11]:} Y. Georgelin and J.C. Wallet, Phys. Lett. A224 (1997) 303.
\item {[12]:} J. K. Jain, Phys. Rev. B41 (1990) 7653; see also third of ref.
[8].
\item {[13]:} T. Allen, Physica 6D (1983) 305. Visibility diagram
have also been used previously in the fractional Hall effect by
J. Fr\"ohlich and E. Thiran from a somehow different viewpoint 
, see Z\"urich preprint ETH-TH/96-22 (Not published).
\item {[14]:} For Farey sequences theory see : G.H. Hardy and
E.M. Wright, an introduction to the theory of numbers,
Clarendon Press, Oxford (1968).
\item {[15]:} See in D. Mumford, Tata Lectures on Theta (vols. I, II, III),
Birkh\"auser, Boston, Basel, Stuttgart (1983); see also in B. Schoeneberg, 
Elliptic Modular Functions, an introduction, Springer-Verlag, Berlin, 
Heidelberg (1974).
\item {[16]:} F. D. M. Haldane and E. H. Rezayi, Phys. Rev. B31 (1985) 2529.
\item {[17]:} G. Christofano, G. Maiella, R. Musto and F. Nicomedi, Phys. Lett.
B262 (1991) 88.
\item {[18]:} R.L. Willet, J.P. Eisenstein, H.L. St\"ormer, D.C. Tsui,
A.C. Gossard and J.H. English, Phys. Rev. Lett. 61 (1988) 997,
and references therein; see also J.K. Jain, talk presented 
at the XXXV Cracow School of Theoretical Physics, Zakopane (Poland), June 1995;
Published in Acta Physica Polonica 26 (1995) 2149.
\item{[19]:} R. Willett et al., Phys. Rev. Lett. 59 (1987) 1776.     
\vfill\eject
\input epsf.tex

\phantom{bof}
\vskip -1cm

\centerline{\epsfbox{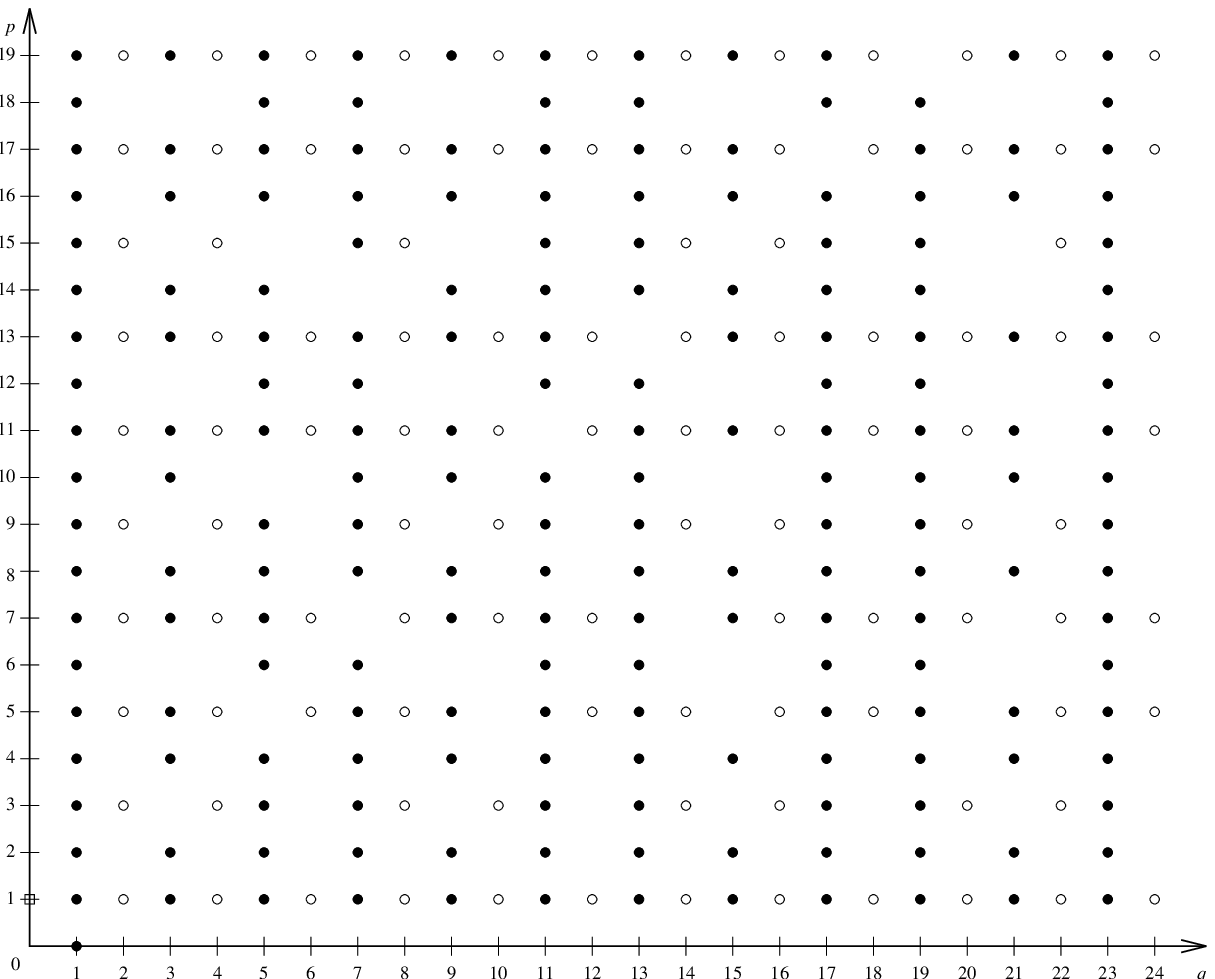}}
\centerline{{\bf Figure 1:} Visibility diagram. Black dots correspond to odd}
\centerline{denominator fractions whereas circles indicate even denominator 
one.}

\centerline{\epsfbox{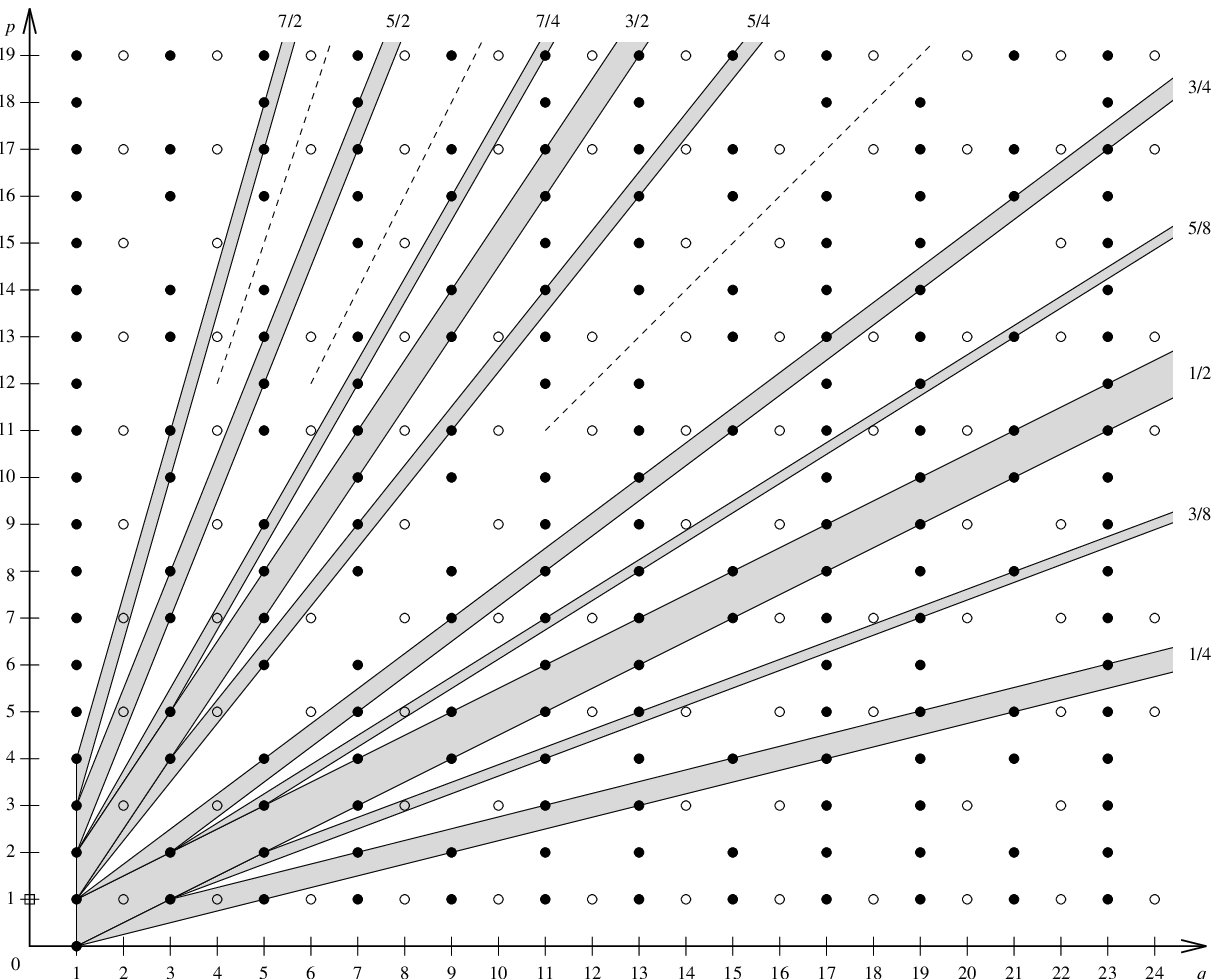}}
\centerline{{\bf Figure 2:} Even denominator stripe structures.}

\centerline{\epsfbox{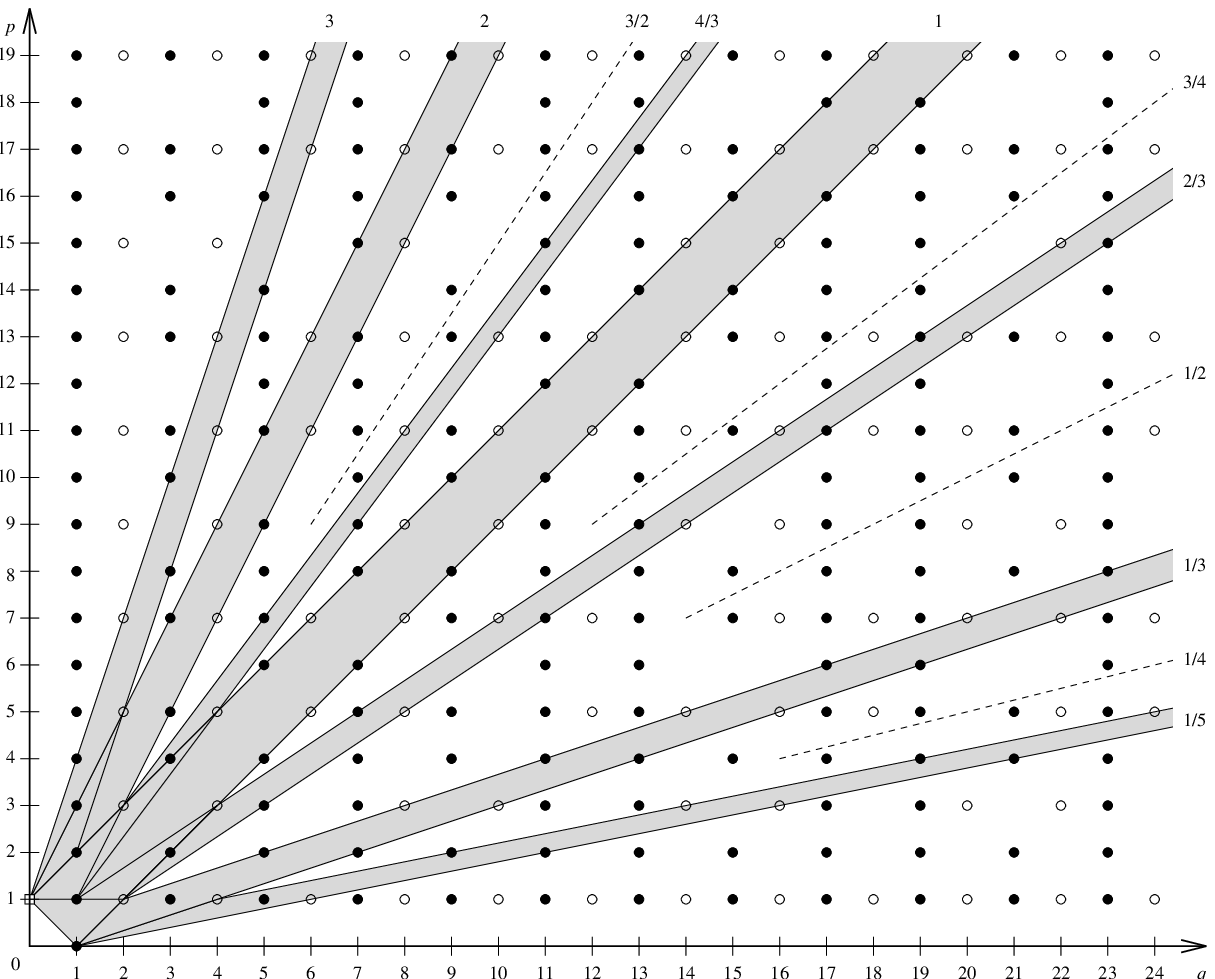}}
\centerline{{\bf Figure 3:} Odd denominator stripe structures.}

\centerline{\epsfbox{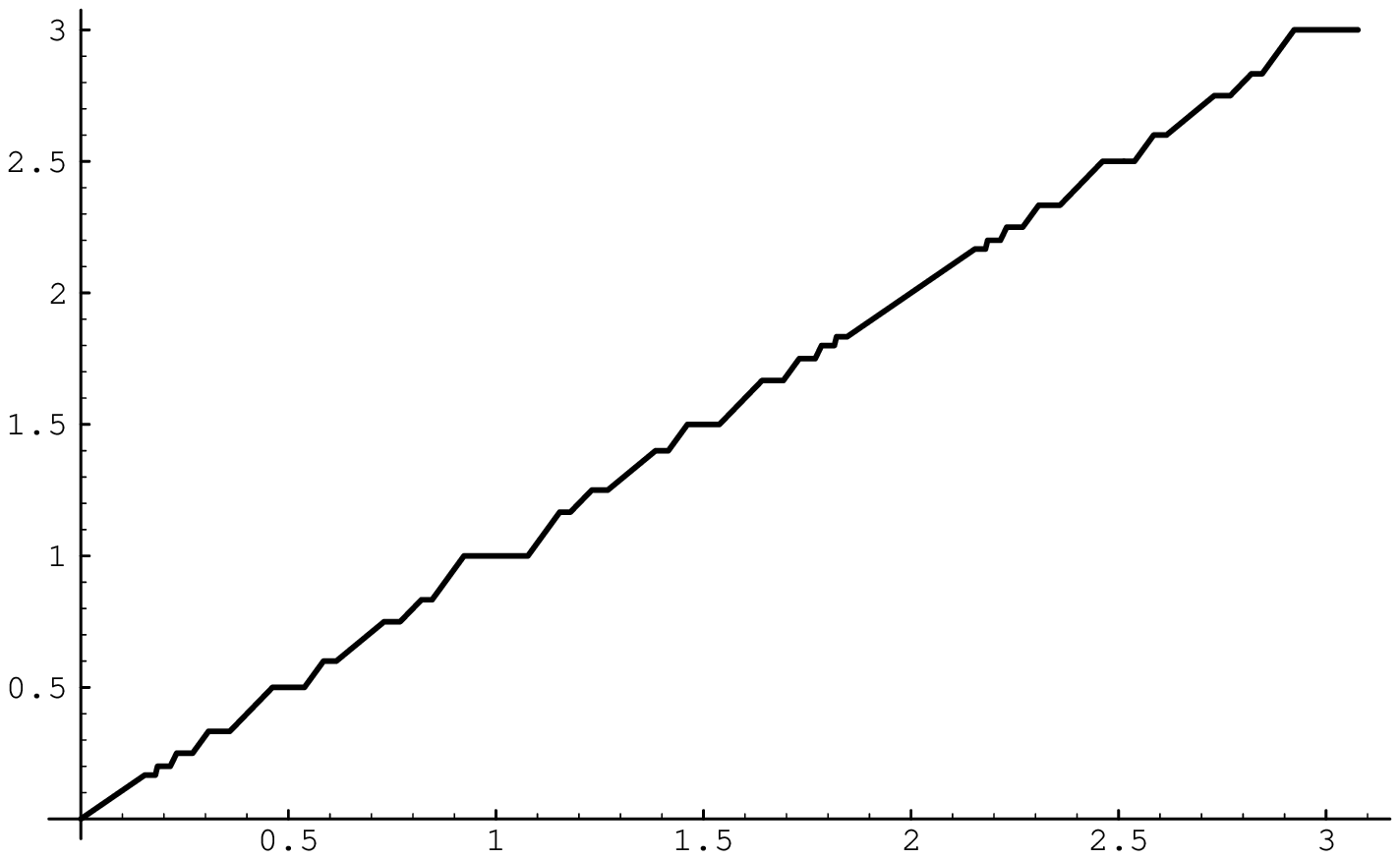}}
\centerline{{\bf Figure 4:} Numerical determination of the transverse
resistivity plot.}
\centerline{ The horizontal (resp. vertical) axis corresponds to filling
factor $\nu$}
\centerline{ (resp. tranverse resistivity in unit ${{h}\over{e^2}}$).}}

\end

{\bf{REFERENCES}}\par

\item {[8]:} \item {[10]:} S. Kivelson, Dung-Hai Lee and Shou-Cheng Zhang, Phys. Rev. B46
(1992) 2223.
\item {[11]:} B. I. Halperin, P. A. Lee and N. Read, Phys. Rev. B47 (1993)
7312.
\item {[12]:} C. A. L\"utken and G. G. Ross, Phys. Rev. B48 (1993) 2500; Phys.
Rev. B48 (1993) 11837; C. A. L\"utken, Nucl. Phys. B396 (1993) 670.
\item {[13]:} P. L\'evay, J. Math. Phys. 36 (1995) 2792.
\item {[14]:} 
\item {[15]:} \item {[16]:} J. E. Avron and R. Seiler, Phys. Rev. Lett. 54 (1985) 259; see
also Q. Niu and D. J. Thouless, Phys. Rev. B35 (1987) 2188.
\item {[17]:} see e.g. R. G. Clark et al., Phys. Rev. Lett. 60 (1988) 1747; 
J. P. Eisenstein, H. L. Stormer, L. N. Pfeiffer and K.
W. West, Phys. Rev. B41 (1990) 7910; H. W. Jiang et al., Phys. Rev. B44 (1991)
8107; L. W. Engel et al., Phys. Rev. B47 (1992) 3418; M. B. Santos et al.,
Phys. Rev. B46 (1992) 13639; R. R. Du et al., Phys. Rev. Lett. 70 (1992) 
2944; H. C. Manoharan and M. Shayegan, Phys. Rev. B50 (1994) 17662.
\item {[18]:} S. V. Kravchenko, W. Mason, J. E. Furneaux and V. M. Pudalov to
be published in Phys. Rev. Lett.
\item {[19]:} see T. M. Apostol, Modular functions and Dirichlet series in
number theory, Springer-Verlag, New-York, Berlin, Heidelberg (1976).
\item {[20]:} E. Fradkin and S. Kivelson, preprint University of
Urbana-Champaign (1996).
\vfill\eject

[4]: J. Keener, Transactions of the American Mathematical Society
261 (1980) 589.\hfill\break

[5]: We are indebted to J. Bellissard to have pointed to us
the similar role played by ours $A^{\lambda}$ and
operators already introduced by D.R. Hofstadter, Phys. Rev.
 B14 (1976) 2239.\hfill\break
[8]: Concretely, the reader may imagine himself or herself
standing at the corner of an infinitely extended plantation
of tall and thin kind of trees (palmtrees), located at the
odd $q$ vertices, mixed with an other little kind (palmitos)
located at even $q$ ones. The configuration we are considering
here can be fully identifyed with what he or her is gazing at.
[10]: See the  results reported in [2].\hfill\break
[11]: In the plantation model, $\Delta _{p/q}$ corresponds
roughly to a region of maximal density of palmtrees
seen behind a given one located at $(q,p)$.\hfill\break
[12]: For a thorough discussion of those concepts
see: 
[13]: See for instance T.M. Apostol, Modular functions
and Dirichlet series in number theory, Springer-Verlag,
New-York, Berlin, Heidelberg (1976).\hfill\break
[14]: As for instance in recent works on F.Q.H.E. and
Rational Conformal Field Theory by M. Flohr, Mod. Phys.
Lett. A11 (1996) 55; F.Q.H.E. and $W_{1+ \infty}$ by
A. Capelli, G.R. Zemba, Firenze preprint DFF 249/5/96
(hep-th/9605127) or the more classical Chern-Simons
approach by E. Fradkin S. Kivelson, University of
Urbana-Champaign preprint (cond-mat/96003156). \hfill\break
[15]: See text book like B. Doubrovine, S. Novikov,
A. Fomenko, g\'eom\'etrie contemporaine, m\'ethodes et applications
Mir, Moscou (French translation of a Russian original,
English and German translations also available).\hfill\break
[16]: A good introduction for that topic is
A.P. Balachandran,
G. Marmo, B.S. Skagerstam and A. Stern, Classical
Topology and Quantum States, World Scientific,
Singapore, 1991.\hfill\break
[18]: S.V. Kravchenko, W. Mason, J.E. Furneaux and V.M.
Pudalov, Phys. Rev. Lett. 75 (1995) 910.
[19]: H.W. Jiang et al., Phys. Rev. B44 (1991) 8107; M.B. Santos et al., Phys.
Rev. B46 (1992) 13639; P.J. Rodgers et al., J. Phys.:Condens. Matter 5, L449
(1993); H.C. Manoharan and M. Shayegan, Phys. Rev. B50 (1994) 17662.
[20]: H.W. Jiang, C.E. Johnson, K.L. Wang and S.T. Hannahs, Phsy. Rev. Lett. 71
(1993) 1439; T. Wang, K.P. Clark, G.F. Spencer, A.M. Marck and W.P. Kirk, Phys.
Rev. Lett. 72 (1994) 709; R.J.F. Hughes et al., J. Phys.: Condens. Matter 6
(1994) 4763.

\overfullrule=0mm
\input epsf.tex
\vfill\eject

\phantom{bof}

\centerline{\epsfxsize=456pt \epsfbox{fig1.eps}}
{\bf {Figure 1}}: The visibility diagram: Even denominator fractional
states $(\circ )$; Hall fluid states $(\bullet )$; the insulator state
$(1,0)$; the vertex $(0,1)$ is indicated by an empty square.
Separating (dashed)
lines between the three principal stripes $1/2$, $3/2$,
$5/2$ and their descendants are indicated. Up to
now well established states are included in the grey zone.\par
\vfill\eject
\phantom{bof}

\centerline{\epsfxsize=456pt \epsfbox{fig2.eps}}
{\bf {Figure 2}}: Few metallic stripes are indicated in grey.
The segments $I2 = I_2$, $I1 = I_1$ and $I23 = I_{2/3}$ are also
indicated.\par

\vfill\eject
\phantom{bof}

\centerline{\epsfxsize=316pt \epsfbox{fig3.eps}}
{\bf {Figure 3}}:  Branching structure of the metallic stripe $1/2$ and
its descendants.\par

\vskip 1.5 truecm
\centerline{\epsfxsize=206pt \epsfbox{fig4.eps}}
\vskip 0.5 truecm
{\bf {Figure 4}}: Construction of the surface $\Delta _{p/q}$
attributed to the state $p/q$, with $p\ne 0$.\par

\vfill\eject
\phantom{bof}

\centerline{\epsfxsize=232pt \epsfbox{fig5.eps}}
\vskip 0.5 truecm
{\bf {Figure 5}}: The bretzel $Br$ with its two generating cycles,
 homotopically equivalent to
the Hall sample we consider here (fig.~1 in [1]).\par

\vskip 1.5 truecm
\centerline{\epsfxsize=265pt \epsfbox{fig6.eps}}
{\bf {Figure 6}}: The Cayley tree of order $4$, the covering space of
the bretzel $Br$.

\end